\documentclass[aps,preprint,showkeys,showpacs,amsmath,amssymb,eqnarray,array]{revtex4}
\usepackage{anysize}
\marginsize{1.9cm}{1.9cm}{1.55cm}{1.55cm}
\usepackage{graphicx}
\usepackage{dcolumn}
\usepackage{bm}
\usepackage{setspace}

\begin{document}

\title{WHY A MAGNETIZED QUANTUM WIRE CAN ACT AS AN OPTICAL AMPLIFIER: A SHORT SURVEY\\}
\author{Manvir S. Kushwaha}
\address
{Department of Physics and Astronomy, Rice University, P.O. Box 1892, Houston, TX 77251, USA}

{\centerline {\em A bend in the road is not the end of the road... unless you fail to make the turn.}}
{\centerline {\hspace{11.9cm} --- Unknown}}

\date{\today}
\begin{abstract}

This article reviews the fundamental issues associated with the magnetoplasmon excitations investigated
in a semiconducting quantum wire characterized by a harmonic confining potential and subjected to an
applied (perpendicular) magnetic field. We embark on the charge-density excitations in a two-subband
model within the framework of Bohm-Pines's random-phase approximation. The problem involves two length
scales: ${\it l}_0=\sqrt{\hbar/m^*\omega_0}$ and ${\it l}_c=\sqrt{\hbar/m^*\omega_c}$, which
characterize the strengths of the confinement and the magnetic field ($B$). Essentially, we focus on
the device aspects of the intersubband collective (magnetoroton) excitation, which observes a negative
group velocity between maxon and roton. Consequently, it leads to tachyon-like (superluminal) behavior
without one's having to introduce the negative energies. Existence of the negative group velocity is a
clear manifestation of a medium with population inversion brought about due to a metastable state
caused by the magnetic field that satisfies the condition $B> B_{th}$; $B_{th}$ being the threshold
value below which the magnetoroton does not exist. The interest in negative group velocity is based on
anomalous dispersion in a medium with inverted population, so that gain instead of absorption occurs at
the frequencies of interest. A medium with an inverted population has the remarkable ability of
amplifying a small optical signal of definite wavelength, i.e., it can serve as an {\em active} laser
medium. An extensive scrutiny of the gain coefficient suggests an interesting and important application:
the electronic device designed on the basis of such magnetoroton modes can act as an optical amplifier.
Examining the magnetic-field dependence of the life-time of magnetorotons leads us to infer that
relatively smaller magnetic fields are optimal.

\end{abstract}
\keywords{Quantum wires, magnetic fields, magnetorotons, laser amplifiers}
\pacs{73.21.Hb, 73.43.Lp, 73.63.Nm, 78.67.Lt}
\maketitle


\section{A Kind of Introduction}

In these modern times, even for a science pundit to argue, to a point of rational certainty, that science was
pursued long before the advent of pen, paper, and printing is {\em a hard nut to crack}. In prehistoric times,
the knowledge used to be diffused from generation to generation in an oral tradition. The evolution of writing
enabled knowledge to be stored and communicated across generations with much greater fealty. Although empirical
investigations of the natural world have been described since {\em Classical Antiquity}, and scientific methods
have been employed since the {\em Middle Ages}, the dawn of modern science is often traced back to the early
modern period and in particular to the {\em scientific revolution} that took place in the 16th to 17th century.
Scientific methods are considered to be so fundamental to modern science that some consider earlier inquiries
into nature to be pre-scientific.

Science is a body of empirical, theoretical, and practical knowledge about the natural world, produced by the
scientists who stress upon the observation, explanation, and prediction of real world phenomena. Despite the
slow but steady quest for science over the long time span, the English word {\em scientist} is relatively
recent -- first coined by William Whewell in 1833. Previously, people investigating nature called themselves
{\em natural philosophers}. Subsequently, the term {\em physics} was coined in 1850 by Cardinal Newman, who
defined it to be ``that family of sciences which is concerned with the {\em sensible} world, that is the
phenomena which we see, hear, and touch; it is the philosophy of matter". Gradually, physicists have narrowed
their focus over the past century, perhaps because they have realized that the extent of knowledge has been
growing so great and so fast that few, if any, can encompass it all.

The world's renowned thinkers -- from the past and present -- believe that the transformation of the society
has produced, to be sure, many beautiful ruins, but not a better society. The reason for such disappointments
obviously lies in the man-made catastrophe in Hiroshima and Nagasaki in WWII. The physics and particularly the
condensed matter physics that has an exceptionally poignant creation myth is the core behind such a contention. Nevertheless, if there is any subject that can still claim to lie at the heart of knowledge of the natural
world, it is physics. Condensed matter physics is the largest branch of physics, which has the greatest impact
on our daily lives by providing basis for technological innovations.

Condensed matter physics of the past twenty five years is the physics of the semiconducting systems of reduced
dimensions. Specifically, we refer to the exotic physics emerging from the semiconducting quasi-N-dimensional
electron systems (QNDES); with $N$ ($< 3$) being the degree of freedom. These are quantum wells ($N=2$),
quantum wires ($N=1$), and quantum dots ($N=0$) in which the charge carriers exposed to electric
and/or magnetic fields exhibit novel quantal effects that strongly modify their behavior characteristics. The
quantization of energy levels in inversion layers had been anticipated much earlier [1], but 2D nature of the
electron gas when only the lowest electric sub-band is occupied was confirmed a decade later by experiments on
$n$-type inversion layers of ($100$) surface of silicon in the presence of a perpendicular magnetic field [2].
Our insight of why and how behavior of a physical system depends on spatial dimensionality has been enriched
significantly by studies of Q2DES. All the still-lower dimensional systems are known to be the willful
derivatives of the Q2DES.

The current progress in the nano-fabrication technology and the electron lithography and the ability to tailor
potentials and interactions is stimulated by the world-wide quest to develop exotic high-speed, low-power
devices that are small enough, sharp enough, and uniform enough to behave the way theory says they should. The
discovery of quantum Hall effects -- both integral [3] and fractional [4] -- is credited to have spurred the
gigantic research interest in the systems of reduced dimensionality. In the present work, we are concerned
with the theoretical investigation of quantum wires or (more realistically) a quasi-one dimensional
electron gas (Q1DEG) originally proposed by Sakaki in 1980 [5]. For an extensive review of electronic, optical,
and transport phenomena in QNDES, a reader is referred to Ref. 6.

The literature is a live witness that much of the fundamental theoretical understanding of electron dynamics
in one-dimensional (1D) systems have emerged from the work on the Tomonaga-Luttinger liquid model (TLLM) [7-8].
The TLLM makes some of the drastic simplifying assumptions, which allow one to solve the interacting problem
completely. One of the surprising results obtained from the solution of TLLM is that even the smallest
interaction results in the disappearance of Fermi surface, leading to a system which is describable as a
non-Fermi liquid -- in the sense that the elementary excitations are very different from those of the
noninteracting system. Therefore, one would expect that the experimental properties of the semiconducting
quantum wires should be quite different from any of those based on the assumption that a 1D electron gas (1DEG)
is a Fermi liquid. And yet, contrary observations have been rather firmly established [9].

The proposal of the quantum wire structures was motivated by the suggestion [5] that 1D $k$-space restriction
would severely reduce the impurity scattering, thereby substantially enhancing the low-temperature electron
mobilities. As a result, the technological promise that emerges is the route to the faster electronic devices
fabricated out of quantum wires. A tremendous research interest burgeoned in quantum wires owes not only to
their potential applications, but also to the fundamental physics involved. For instance, they have offered
us an excellent (unique) opportunity to study the real 1D fermi gases in a relatively controlled manner.
Theoretical development of transport phenomena in quantum wires has also been the subject of an intense
controversy over whether the system is best describable as Luttinger liquid or as Fermi liquid. This issue
was elegantly resolved by Das Sarma and coworkers [9] who consistently justified the use of Fermi-liquid-like
theories for describing the realistic quantum wires.

The quantum wires lie midway of the quantum rainbow and are known to possess some unique electronic, optical,
and transport properties such as magnetic depopulation [10], electron waveguide [11], quenching of the Hall
effect [12], quantization of conductance [13], negative-energy dispersion [14], magnetoroton excitations [15],
spin-charge separation [16], to name a few. The theoretical investigations on the quantum wires, particularly
associated with the magnetoplasmon excitations, have not been quite consistent with the experimental findings.
For instance, early theoretical works [17-19], had not been, to our knowledge, able to verify the prediction
of the existence of magnetorotons [18] in quantum wires in the presence of an applied perpendicular magnetic
field. Also, our understanding of the inter-Landau level collective excitations in the quantum wires has been
quite limited in the situations where Kohn's theorem [20] is violated.

This review reports on the recent extensive investigations of the charge-density excitations in a magnetized
quantum wire in a two-subband model within the framework of Bohm-Pines' random-phase approximation (RPA) [21].
Our main focus is the intersubband collective (magnetoroton) excitation which changes the sign of its group
velocity twice before merging with the respective single-particle continuum. By definition, a roton is an
elementary excitation whose dispersion relation shows a linear increase from the origin, but exhibits first
a maximum, and then a minimum in energy as the momentum increases. Excitations with momenta in the linear
region are called phonons; those with momenta near the maximum are called maxons; and those with momenta
close to the minimum are called rotons.

A roton mode in 3D superfluid $^4$He was empirically derived within the two-fluid model by Landau [22], and
its reliable theory was developed and refined by Feynman [23]. In 2D electron gas (2DEG), the magnetoroton
minimum was obtained in the fractional quantum Hall effect regime within the framework of single-mode
approximation by Girvin et al. [24]. In Q1DEG, the magnetoroton mode was predicted within the framework of
Hartree-Fock approximation in 1992 [18] and it was soon verified in the resonant Raman scattering experiments
[15]. A rigorous theoretical finding of the magnetoroton (MR) modes in the realistic quantum wires had,
however, been elusive until quite recently [25]. After a brief theoretical background, we illustrate and
discuss a variety of relevant aspects of the MR mode collected from our recent research on the Q1DEG [25-28].
Then we systematically connect the dots in order to bring about the central idea and substantiate the
argument that an electronic device designed on the basis of such MR modes can act as an optical amplifier.

There is one trait that many theoretical physicists share with philosophers. In both cases the interest in
a field of study seems to vary in inverse proportion to how much one must learn to qualify as an expert.
There is an important core of truth to the fact that ``expertness is what survives when what has been
learnt has been forgotten". True learning leading someone to become an expert is a long-term project --
ideally life-long. However, present day learning of any subfield of science, particularly the vastly
growing (condensed matter) physics, motivated to gain mastery is inconceivably parlous for survival in the
existing competitive world of science. Here, we shall try to review briefly what we have been able to
learn from theoretically motivated analytical diagnoses of the magnetoplasmon excitations -- specifically
the intersubband collective [or {\em magnetoroton}] excitations -- in a realistic quantum wire subjected
to a harmonic confining potential and exposed to an applied (perpendicular) magnetic field.  We appeal to
interested readers to digest what they should and what they can and blame the author for the rest.

The rest of the article is organized as follows. In Sec. II, we list some fundamental mathematical formulae
derived for the QNDES in the absence of an applied magnetic field. Generalization of these formulae to
include the magnetic field is quite straightforward. In Sec. III, we present the theoretical framework in
order to derive and discuss the full Dyson equation (that takes proper account of the Coulombic interactions)
and the nonlocal, dynamic, dielectric function utilized for investigating the single-particle and collective (magnetoplasmon) excitations within the framework of RPA. In Sec. IV, we discuss several illustrative
examples, which focus mainly on the (intersubband) magnetoroton excitations computed in a two-subband model
within the full RPA. This includes, e.g., examining the effect of several experimental parameters such as
the magnetic field, the charge density, and the confinement potential. Then we examine the dependence on the
damping of the gain coefficient for the magnetoroton excitations. Lastly, we shed some light on
the magnetic-field-dependence of the life-time of magnetorotons in the realistic quantum wires. Finally, we
conclude our findings and suggest some interesting dimensions worth adding to the problem in Sec. V.

\begin{figure}[htbp]
\includegraphics*[width=10cm,height=7cm]{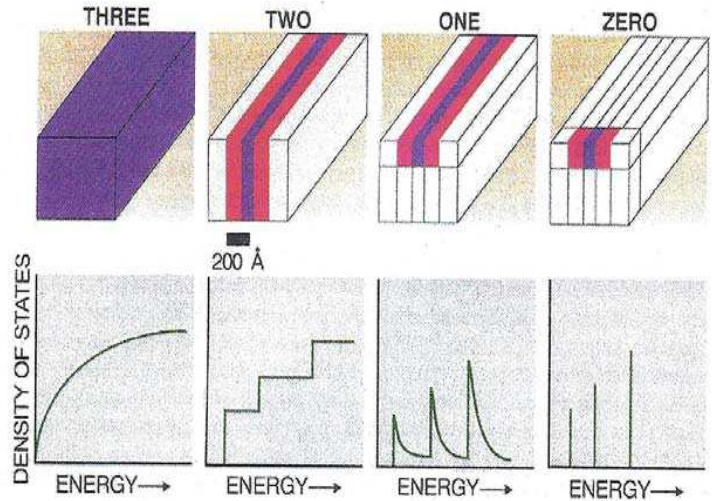}
\caption{(Color online) Schematics of the quantum confinement (top panel) for achieving quasi-n-dimensional
($n = 3; 2; 1;$ and 0) systems. The regions in purple refer to the desired quasi-n-dimensional systems, whose dimensionality can be reduced by sandwiching it between two layers of another material with larger band gap.
The confinement changes the density of states (DOS) [bottom panel], or specific energy levels, that will be
filled by incoming electrons. The current conducted by a quantum-well device peaks when the energy level of
the incoming electrons matches (or is in resonance with) the energy level of the quantum well. (After M.S.
Kushwaha, Ref. 6).}
\label{fig1}
\end{figure}

\section{Fundamentals of QNDES}

Here we would like to list a few fundamental mathematical expressions derived for the QNDES [see Fig. 1]
in the absence of an applied magnetic field. These relations are very useful in order to understand
sophisticated results obtained for the electronic, optical, and transport phenomena in the QNDES. They
include, e.g., Fourier transform of the (binary) Coulombic interactions, density of states, Fermi energy,
plasma frequency, to mention a few. The Coulomb interaction in the wave vector space is given by the
appropriate quasi-$N$-dimensional Fourier transform of the Coulomb interaction
$V_{QN}$(${\boldsymbol r}$)$=e^2/\epsilon_b {\boldsymbol r}$ as
\begin{subequations}
\begin{align}
V_{Q3}(q)&=\frac{4\pi e^2}{\epsilon_b q^2}\, ,\\
V_{Q2}(q; z, z')&=\frac{2\pi e^2}{\epsilon_b \, q}\,e^{-q\mid z-z'\mid}\, ,\\
V_{Q1}(q; x, x')&=\frac{2e^2}{\epsilon_b}\, K_0(q\mid x-x'\mid)\, ,
\end{align}
\end{subequations}
where $\epsilon_b$ is the background dielectric constant of the medium in which QNDES is embedded. The density
of states for the corresponding systems is given by
\begin{subequations}
\begin{align}
D_{Q3}(\epsilon)&=2\,\frac{1}{4\pi^2}\, \big(\frac{2\,m^*}{\hbar^2}\big)^{3/2}\, \epsilon^{1/2}\, ,\\
D_{Q2}(\epsilon)&=2\,\frac{1}{4\pi}\,\big(\frac{2\,m^*}{\hbar^2}\big)\,\sum_n\,\theta
                                                      \big(\epsilon - \epsilon_n\big)\, ,\\
D_{Q1}(\epsilon)&=2\,\frac{1}{2\pi}\,\Big (\frac{2 m^*}{\hbar^2}\Big )^{1/2}\,\sum_{n}\,
              \big (\epsilon - \epsilon_n \big)^{-1/2}\, \theta \big(\epsilon - \epsilon_n\big)\, ,
\end{align}
\end{subequations}
where $\epsilon_n$ is the eigenenergy of the respective $n$th subband and $\theta$($x$) is the well-known
Heaviside unit step function. The factor of 2 accounts for the spin degeneracy.  Next, the Fermi energy for
the corresponding systems is specified by
\begin{subequations}
\begin{align}
n_{Q3}&=\frac{1}{3\pi^2}\, \Big(\frac{2\,m^*}{\hbar^2}\Big)^{3/2}\, \epsilon^{3/2}_F\, ,\\
\sqrt{n_{Q2}}&=\Big (\frac{m^*}{\pi\,\hbar^2}\Big )^{1/2}\,\sum_{n}\,
     \big (\epsilon_F - \epsilon_n \big )^{1/2}\, \theta (\epsilon_F - \epsilon_n)\, ,\\
n_{Q1}&=\frac{2}{\pi}\,\Big (\frac{2 m^*}{\hbar^2}\Big )^{1/2}\,
\sum_{n}\, \big (\epsilon_F - \epsilon_n\big )^{1/2}\, \theta (\epsilon_F - \epsilon_n)\, ,
\end{align}
\end{subequations}
The Fermi wave vector $k_F$ and the particle density $n_{QN}$ are related such that $n_{Q3}=k_F^3/(3\pi^2)$,
$n_{Q2}=k_F^2/(2\pi)$, and $n_{Q1}=2k_F/\pi$ in the corresponding systems. The long-wavelength screened plasma
frequency $\omega_p^{QN}$ in the QNDES is deduced as
\begin{subequations}
\begin{align}
\omega_p^{Q3}&=\sqrt{\frac{4\pi\,n_{Q3}\,e^2}{\epsilon_b\, m^*}}+O(q^2)\, ,\\
\omega_p^{Q2}&=\sqrt{\frac{2\pi\,n_{Q2}\,e^2}{\epsilon_b\, m^*}}\,q^{1/2}+ O(q^{3/2})\, ,\\
\omega_p^{Q1}&=\sqrt{\frac{2\pi\,n_{Q1}\,e^2}{\epsilon_b\, m^*}}\,q\,\sqrt{\mid\ln(qt)\mid} +O(q^3)\, ,
\end{align}
\end{subequations}
where $n_{QN}$ is carrier density per unit N-dimensional volume. The purpose of deriving Eqs. (4) is simply to
demonstrate that the long-wavelength plasmon frequency for a harmonically confined systems is purely classical
since $\hbar$ does not appear in the leading term of Eqs. (4) in any dimension. The second order dispersion
correction term in Eqs. (4) is fully quantal since $\hbar$ shows up explicitly and is affected by interaction
corrections (both self-energy and vertex corrections to the irreducible polarizability) [29]. The symbol $e$ is
the elementary electronic charge and $m^*$ the electron effective mass. The rest of the symbols are defined in
what follows. The generalization of Eqs. (1)$-$(4) to include the applied magnetic field is not so difficult.


\section{Methodological Framework}

We consider a Q1DEG in the $x-y$ plane with a harmonic confining potential $V(x)=\frac{1}{2}m^*\omega_0^2 x^2$
oriented along the $x$ direction and a magnetic field $B$ applied along the $z$ direction in the Landau gauge
[$\bf {A}=(0,Bx,0)$] (see, e.g., Fig. 2). Here $\omega_0$ is the characteristic frequency of the harmonic
oscillator and $m^*$ the electron effective mass of the system. The resultant system is a realistic quantum
wire with free propagation along the $y$ direction and the {\em magnetoelectric} quantization along the $x$. A
realistic quantum wire as defined above is characterized by the eigenfunction
\begin{equation}
\psi_{n}({\boldsymbol r})=\frac{1}{\sqrt{L_y}}\, e^{ik_y y}\,\phi_n(x+x_c),
\end{equation}
where ${\boldsymbol r} \equiv (x, y)$ is a 2D vector in the {\em direct} space and $\phi_n(x+x_c)$ is the
Hermite function, and the eigenenergy
\begin{equation}
\epsilon_{nk_y} = (n + \frac{1}{2})\,\hbar\tilde{\omega} + \frac{\hbar^2 k_y^2}{2 m_r},
\end{equation}
where $L_y$, $n$, $x_c=k_y ({\it l}_d^4/{\it l}_c^2)$, ${\it l}_c=\sqrt{\hbar/(m^*\omega_c)}$,
${\it l}_d=\sqrt{\hbar/(m^*\tilde{\omega})}$, $m_r=m^*(\tilde{\omega}^2/\omega_0^2)$, are, respectively, the
normalization length, the hybrid magnetoelectric subband (HMES) index, the center of the cyclotron orbit
with radius ${\it l}_d$, the magnetic length, the effective magnetic length, and the renormalized effective
mass. Here the hybrid frequency $\tilde{\omega}=\sqrt{\omega_c^2+\omega_0^2}$. The effective magnetic length
${\it l}_d$ refers to the typical width of the wave function and reduces to the magnetic length ${\it l}_c$
if the confining potential is zero (i.e., if $\omega_0=0$). In the limit of a strong magnetic field,
the renormalized mass $m_r$ becomes infinite and the system undergoes a cross-over to the 2DES and hence the
Landau degeneracy is recovered. In what follows we choose to drop the subscript y on all the quantities for
the sake of brevity. Also, it is interesting to start with the single-particle density-density correlation
function (DDCF) given by [6]
\begin{equation}
\chi^{0} ({\boldsymbol r},{\boldsymbol r'}; \omega)=\sum_{ij}\, \Lambda_{ij}\,\,
\psi^*_i ({\boldsymbol r'})\,\psi_j ({\boldsymbol r'})\,
\psi^*_j ({\boldsymbol r})\,\psi_i ({\boldsymbol r})\, ,
\end{equation}
where the subindex $i,j\equiv k,n$ and the substitution $\Lambda_{ij}$ is defined by
\begin{equation}
\Lambda_{ij}= 2\, \frac{f(\epsilon_i)-f(\epsilon_j)}{\epsilon_i-\epsilon_j+\hbar\omega^+} \, ,
\end{equation}
Here $f(x)$ is the familiar Fermi distribution function. $\omega^+=\omega+i\gamma$ and small but nonzero
$\gamma$ refers to the adiabatic switching of the Coulombic interactions in the remote past. The factor
of $2$ accounts for the spin degeneracy. Next, we recall the Kubo's correlation function to express the
induced particle density
\begin{align}
n_{in}({\boldsymbol r}, \omega)
&= \int d{\boldsymbol r}'\, \chi ({\boldsymbol r},{\boldsymbol r'}; \omega)\,V_{ex}({\boldsymbol r'}, \omega) \nonumber\\
&= \int d{\boldsymbol r}'\, \chi^{0} ({\boldsymbol r},{\boldsymbol r'}; \omega)\,V({\boldsymbol r'}, \omega)
\end{align}
Here $V(...)=V_{ex}(...)+V_{in}(...)$ is the total potential, with subscripts ex (in) referring to the
external (induced) potential. Let us mention {\em once and for all} that the term potential, in fact,
refers to potential energy. Here $\chi (...)$ [$\chi^0$(...)] is the total [single-particle] DDCF.
Eqs. (9) are Fourier-transformed with respect to time factor $t$. Further, the induced potential in terms
of the induced particle density is given by
\begin{equation}
V_{in}({\boldsymbol r}, \omega)=\int d{\boldsymbol r}'\, V_{ee}
                     ({\boldsymbol r}, {\boldsymbol r}')\, n_{in}(x', \omega) ,
\end{equation}
where $V_{ee}(...)$ refers to binary Coulomb interactions and its 1D Fourier transform is defined by
\begin{equation}
V_{ee}(q; x, x')=\frac{2e^2}{\epsilon_b}\, K_0(q\mid x-x'\mid)
\end{equation}
where $K_0 (x)$ is the zeroeth-order modified Bessel function of the second kind, which diverges as $-\ln (x)$
when $x\rightarrow 0$. It is now quite straightforward to prove, from Eqs. (9) and (10), that $\chi(...)$ and $\chi^0(...)$ are, in fact, correlated through the famous Dyson equation
\begin{equation}
\chi (x, x')=\chi^0 (x, x') +\int dx'' \int dx''' \, \chi^0 (x, x'')\, V_{ee}(x'', x''')\, \chi (x''', x')\, ,
\end{equation}
where $\chi$(...) and $\chi^0$(...) are also Fourier transformed with respect to spatial coordinate $y$
and ($q, \omega$)-dependence is suppressed for the sake of brevity. The Dyson equation characteristically
represents the quantum system and is known to serve useful purpose for investigating various electronic,
optical, and transport properties such as, e.g., the optical (or magneto-optical) absorption in the system.
A simple mathematical manipulation of Eqs. (9) and (10), employing the condition of self-consistency:
$V(...)=V_{ex}(...)+V_{in}(...)$, and imposing the setup of self-sustaining plasma oscillations yields the
generalized nonlocal, dynamic, dielectric function expressed as
\begin{equation}
\epsilon_{mm'nn'}(q, \omega)=\delta_{mn}\delta_{m'n'} - \Pi_{nn'}(q, \omega)\,F_{mm'nn'}(q)\, ,
\end{equation}
where $q=k'-k$ is the momentum transfer,
\begin{equation}
\Pi_{nn'}(q, \omega)
=\frac{1}{L}\,\sum_k \,\Lambda_{nn'}(...)
=\frac{2}{L}\,\sum_k \,\frac{f(\epsilon_{nk})-f(\epsilon_{n'k'})} {\epsilon_{nk}-\epsilon_{n'k'}+\hbar\omega^+}
\end{equation}
is, generally, termed as the polarizability function, and
\begin{equation}
F_{mm'nn'}(q)=\frac{2\,e^2}{\epsilon_b}\, \int dx \int dx' \,
\phi^*_{m}(x)\,\phi_{m'}(x)\, K_0(q\mid x-x'\mid)\,\phi^*_{n'}(x')\,\phi_{n}(x')\, ,
\end{equation}
refers to the matrix elements of the Coulombic interactions. This implies that the excitation spectrum should
be computed through searching the zeros of the determinant of the matrix $\tilde{\epsilon}$ ($q, \omega$)
[i.e., $\mid \tilde{\epsilon}(q, \omega) \mid =0$]. In general, our aim is to investigate the single-particle
and collective excitations in a magnetized quantum wire within the framework of Bohm-Pines' RPA in a
two-subband model, with only the lowest one assumed to be occupied. This demands limiting the number of
subbands involved in the problem; otherwise, the general matrix, with elements $\epsilon_{mm'nn'}(...)$,
is an $\infty \times \infty$ matrix and naturally impossible to solve.
Limiting ourselves to a two-subband model is a reasonable choice for such low-density, low-dimensional systems,
particularly at low temperatures where most of the experiments are performed. Consequently,
$\tilde{\epsilon}(q, \omega)$ is a $4 \times 4$ matrix defined by
\begin{equation}
\tilde{\epsilon}(q,\omega)=
\begin{bmatrix}
1-\Pi_{11}\,F_{1111} \ & \ -\Pi_{11}\,F_{1112} \ & \ -\Pi_{11}\,F_{1121} \ & \ -\Pi_{11}\,F_{1122} \\
-\Pi_{12}\,F_{1211} \ & \ 1-\Pi_{12}\,F_{1212} \ & \ -\Pi_{12}\,F_{1221} \ & \ -\Pi_{12}\,F_{1222} \\
-\Pi_{21}\,F_{2111} \ & \ -\Pi_{21}\,F_{2112} \ & \ 1-\Pi_{21}\,F_{2121} \ & \ -\Pi_{21}\,F_{2122} \\
-\Pi_{22}\,F_{2211} \ & \ -\Pi_{22}\,F_{2212} \ & \ -\Pi_{22}\,F_{2221} \ & \ 1-\Pi_{22}\,F_{2222}
\end{bmatrix}\, .
\end{equation}
Since $\Pi_{22}=0$ because the second subband is empty, Eq. (16) simplifies to make the computation a little
less cumbersome. In addition, it is becoming fairly known that, for a symmetric potential (as is the case here),
the Fourier transform of the Coulombic interactions $F_{ijkl}(q)$ is strictly zero provided that $i+j+k+l=$
an odd number [6]. The reason is that the matching eigenfunction is either symmetric or antisymmetric under
space reflection. This implies further simplification of Eq. (16) due to the nature of the confining potential.

\begin{figure}[htbp]
\includegraphics*[width=8cm,height=9cm]{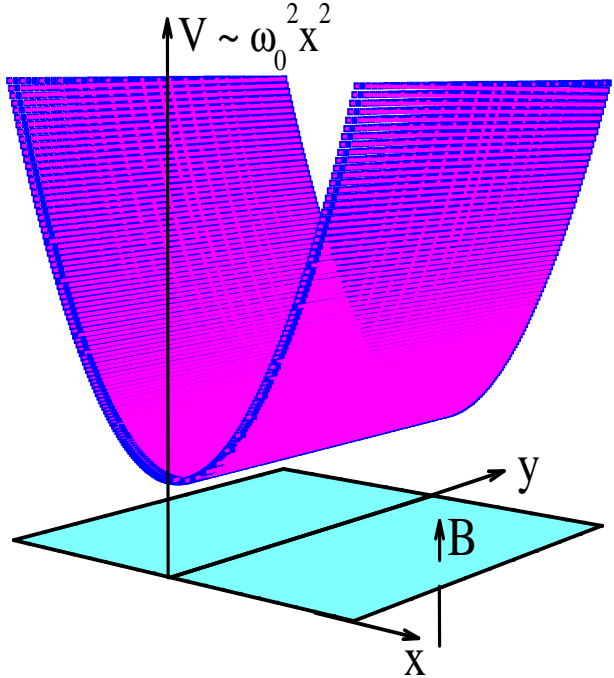}
\caption{(Color online) The model quantum wire [or Q-1DEG for better and broader range of physical
understanding] investigated in this work. The magnetic field $B \parallel {\hat z}$ axis in the Landau
gauge [$\bf {A}=(0,Bx,0)$] and the (harmonic) confining potential imposed on an otherwise 2DEG is
defined as $V(x)=\frac{1}{2}m^*\omega_0^2 x^2$. The resultant system is a realistic quantum wire with
free propagation along $y$ direction and the {\em magnetoelectric} quantization along $x$. (After M.S.
Kushwaha, Ref. 25).}
\label{fig2}
\end{figure}

\section{Illustrative Examples}

For the illustrative numerical examples, we have been focusing  on the {\em narrow} channels of the narrow-gap
In$_{1-x}$Ga$_x$As system. The material parameters used are: effective mass $m^*=0.042 m_{_0}$, the background
dielectric constant $\epsilon_{_b}=13.9$, 1D charge density $n_{1D}=1.0\times 10^{6}$ cm$^{-1}$, confinement
energy $\hbar\omega_0=2.0$ meV, and the effective confinement width of the harmonic potential well, estimated
from the extent of the Hermite function, $w_{eff}=40.19$ nm. Notice that the Fermi energy $\epsilon_F$ varies
in the case where the charge density ($n_{1D}$), the magnetic field ($B$), or the confining potential
($\hbar \omega_{0}$) is varied. Thus we set out on the extensive investigation associated with the
magnetoplasmon excitations in a Q1DEG subjected to a perpendicular magnetic field $B$ at T=0 K [25-28].

The full-fledged MR mode was observed in the similar physical conditions in the magnetized quantum wires made up
of narrow-gap In$_{1-x}$Ga$_x$As systems in a rather different context [25]. Subsequently, we extensively studied
the dependence of its propagation characteristics on the charge-density, confinement potential, and magnetic
field [26]. Since the existence of this MR mode is exclusively attributed to the applied (perpendicular) magnetic
field, we also scrutinized its group velocity for numerous values of the magnetic field in order to judge when
and where this otherwise regular intersubband mode starts taking up the magnetoroton character. It was found that
there really is a minimum (threshold) value of the magnetic field ($B_{th}$) below which this MR does not exist
and a regular intersubband magnetoplasmon survives. The $B_{th}$, for the present system, is defined as $B_{th}
\simeq 1.0$ T. There it was also established that both maxon and roton are the higher density of excitation states.

Later, we also embarked on investigating the inverse dielectric functions (IDF) for this system under the similar
physical conditions [27]. The motivation there was not solely to reaffirm the fact that the poles of the IDF and
the zeros of the dielectric function (DF) yield exactly identical excitation spectrum, but also to pinpoint
the remarkable advantage of the former over the latter. For instance, the imaginary (real) part of the IDF sets
to furnish a significant measure of the longitudinal (Hall) resistance in the system. Moreover, the quantity
Im [$\epsilon^{-1} (q,\omega)$] also implicitly provides the reasonable estimates of the inelastic electron
(or Raman) scattering cross-section $S$($q$) for a given system. For the details of the formalism of the IDFs for
the quasi-n dimensional systems, a reader is referred to Ref. 30.

The main concern here is the MR mode between maxon and roton where it exhibits negative group velocity (NGV). We
believe that the occurrence of NGV is quite unusual and hence must have some dramatic consequences. It turns out
that the NGV matters -- in the phenomena such as techyon-like (superluminal) behavior [31], anomalous dispersion
in the gain medium, a state with inverted population (likely) characterized by negative temperature, a medium
having the ability of amplifying a small optical signal and hence serving as an {\em active} laser medium, ...etc.
We focused on the latter and chose to compute the gain coefficient $\alpha$($\omega$) in order to materialize the
notion of a quantum wire acting as an optical amplifier [28].

In the classical electrodynamics, it makes sense to express $\alpha$($\omega$) in terms of Im [$\chi^0(\omega)$],
with $\chi$ being the susceptibility. But the story takes a different turn when it comes to the quantum systems,
as is the case here. Despite the fact that it is $\chi^0$ that, generally, gives rise to (mathematically)
complex nature of the DF (or the IDF), we must recognize that $\chi^0$ contains only the single-particle response,
whereas the IDF can provide both single-particle and collective responses. Since our concern here is the MR which
happens to be the intersubband collective excitation, we ought to search $\alpha$($\omega$) in terms of Im [$\epsilon^{-1}_{inter} (q, \omega)$] rather than Im [$\chi^0_{inter} (q, \omega)$].

\begin{figure}[htbp]
\includegraphics*[width=8cm,height=9cm]{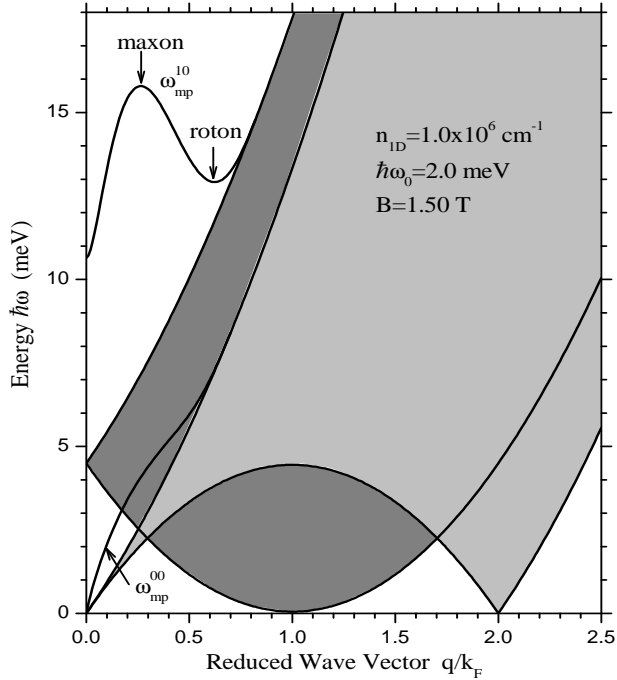}
\caption{(Color Online) The full magnetoplasmon dispersion plotted as energy $\hbar \omega$ vs reduced wave
vector $q/k_F$. The light (dark) shaded region refers, respectively, to the intrasubband (intersubband)
single-particle  excitations. The bold curve marked as $\omega^{00}_{mp}$ ($\omega^{10}_{mp}$) is the
intrasubband (intersubband) collective (magnetoplasmon) excitations. The other parameters are as given inside
the picture. (After M.S. Kushwaha, Ref. 25).}
\label{fig3}
\end{figure}

Figure 3 illustrates the full magnetoplasmon dispersion for the  Q1DEG plotted as energy $\hbar \omega$ vs
reduced wave vector $q/k_F$, for the given values of $n_{1d}$, $\hbar\omega_0$, and $B$. This light (dark)
shaded region refers, respectively, to the intrasubband (intersubband) single-particle excitations (SPE).
The bold solid curve marked as $\omega^{00}_{mp}$ ($\omega^{10}_{mp}$) is the intrasubband (intersubband)
collective (magnetoplasmon) excitations (CME). The intrasubband CME start from the origin and merges
with the upper edge of the intrasubband SPE at ($q_y/k_F=0.66$, $\hbar\omega=7.68$ meV)  and thereafter
ceases to exist as a bonafide, long-lived CME. The intersubband CME starts at ($q_y/k_f=0$,
$\hbar\omega=10.65$ meV), attains a maximum at ($q_y/k_f=0.28$, $\hbar\omega=15.80$ meV), reaches its
minimum at ($q_y/k_f=0.63$, $\hbar\omega=12.93$ meV), and then rises up to merge with the upper edge of
the intersubband SPE at ($q_y/k_f=0.82$, $\hbar\omega=14.63$ meV). It is a simple matter to check
(analytically) why the energy of the lower branch of the intrasubband SPE goes to zero at $q_y=2k_F$, why
the lower branch of the intersubband SPE observes its {\em minimum} ({\em not zero}) at $q_y=k_F$, and why
the intersubband SPE starts at the subband spacing ($\hbar\tilde{\omega}=4.4915$ meV) at the origin. The
most interesting aspect of this excitation spectrum is the existence of this intersubband CME (referred
to as the magnetoroton). Notice how the magnetoroton mode changes its group velocity twice before merging
with the respective SPE. At $q_y=0$, the energy difference between the intersubband CME and SPE is a
manifestation of the many-body effects such as depolarization and excitonic shifts [6]. The knowledge
of the magnetoplasmon dispersion (Figure 3) is of paramount importance to the understanding of the
results in Figs. 9 and 10 (see below).

\begin{figure}[htbp]
\includegraphics*[width=8cm,height=9cm]{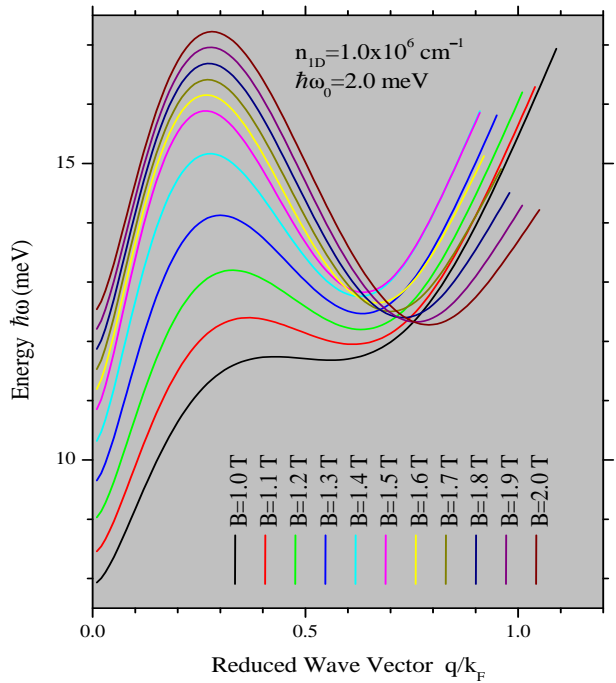}
\caption{(Color Online) MR dispersion plotted as energy $\hbar\omega$ vs reduced wave vector $q/k_F$
for various values of magnetic field (B), for the given values of $n_{1d}$ and $\hbar\omega_0$.
(After M.S. Kushwaha, Ref. 26).}
\label{fig4}
\end{figure}

The full magnetoplasmon spectrum within a two-subband model using the RPA was first illustrated in Fig. 9
in Ref. 25. We call attention to the most curious part of that excitation spectrum --- the existence of
the intersubband collective (magnetoplasmon) excitation (CME) [henceforth referred to as the magnetoroton],
which changes the sign of its group velocity twice before merging with the respective single-particle
excitation (SPE). The interesting thing about it's very occurrence in Q1DEG [25] leads us to infer that
you don't have to overplay with the theory, as was done in Refs. [17] and [19], which both missed to
observe the MR mode. The said magnetoroton excitation dispersion in the energy--wave-vector space is
illustrated in Fig. 4. Each MR mode corresponds to a given magnetic field and for the fixed values of the
charge density ($n_{1D}$) and the confining potential ($\hbar \omega_{0}$). The important feature noticeable
from Fig. 4 is that as the magnetic field ($B$) is increased, the maxon maximum shifts to the higher energy
whereas the roton minimum first observes an increase and then (after a certain value of $B$, here $B=1.5$T)
a decrease in energy.

\begin{figure}[htbp]
\includegraphics*[width=8cm,height=9cm]{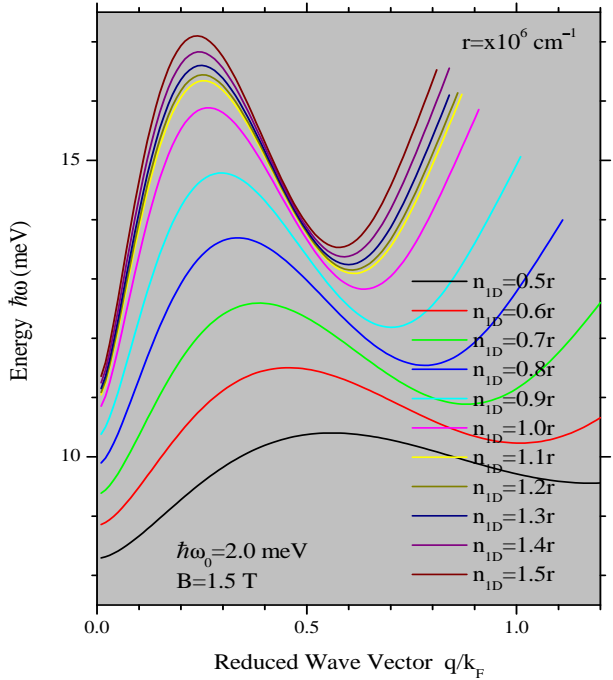}
\caption{(Color Online) MR dispersion plotted as energy $\hbar\omega$ vs reduced wave vector $q/k_F$ for
various values of charge density ($n_{1D}$), for the given values of $B$ and $\hbar\omega_0$.
(After M.S. Kushwaha, Ref. 26).}
\label{fig5}
\end{figure}

Figure 5 shows the MR dispersion for various values of the charge density and for the given values of the
magnetic field and the confinement energy. Unlike the variation of $B$ (Fig. 4), Fig. 5 makes it evident
that there is a systematic trend in the behavior characteristics of the MR as the charge density varies.
To be specific, both the maxon maximum and roton minimum gradually shift to the higher energy (and longer
wavelength) with increasing charge density. It is noteworthy to see how the roton minimum becomes deeper
with increasing $n_{1D}$.

\begin{figure}[htbp]
\includegraphics*[width=8cm,height=9cm]{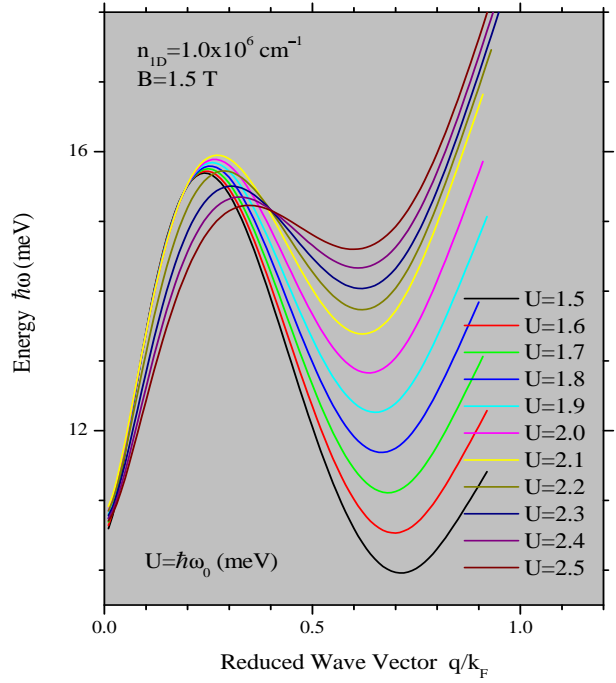}
\caption{(Color Online) MR dispersion plotted as energy $\hbar\omega$ vs reduced wave vector $q/k_F$ for
various values of confinement energy ($\hbar\omega_0$), for the given values of $n_{1d}$ and $B$.
(After M.S. Kushwaha, Ref. 26).}
\label{fig6}
\end{figure}

Figure 6 depicts the MR dispersion for various values of confinement energy and for the given values of
the charge density and the magnetic field. It is interesting to notice here that the roton minimum
observes a systematic shift to the higher energy (and longer wavelength) with increasing
confinement potential. On the other hand, the maxon maximum first observes an increase and then (after
a certain value of $\hbar\omega_0$, here $\hbar\omega_0=2.1$ meV) a decrease in energy with increasing
confinement potential. However, the maxon maximum always tends to attain a shorter wavelength
with increasing $\hbar\omega_0$. A distinctive feature of Fig. 6 (as compared to Figs. 4 and 5) is that
the MR mode here starts (at the origin) within a relatively narrower energy range even though
the confinement potential varies. There is a common feature observed in all Figs. 4, 5, and 6 (not shown
here): every MR mode in the short wavelength limit merges with the upper branch of the respective
(intersubband) SPE.

\begin{figure}[htbp]
\includegraphics*[width=8cm,height=9cm]{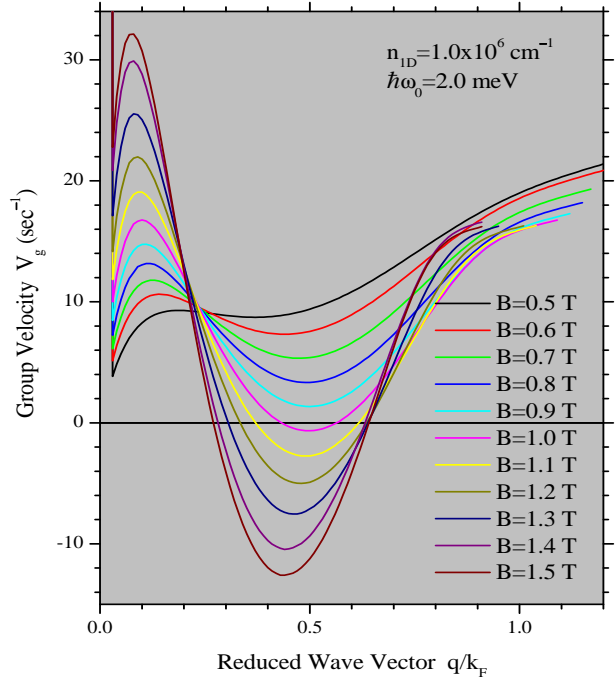}
\caption{(Color Online) The group velocities of the MR excitations plotted in Fig. 1 as a function of
the reduced wave vector $q/k_F$ for several values of $B$ and for the given values of $n_{1d}$ and
$\hbar\omega_0$. (After M.S. Kushwaha, Ref. 26).}
\label{fig7}
\end{figure}

Figure 7 illustrates the group velocities of the MR excitations (plotted in Fig. 4) as a function of the
reduced wave vector $q/k_F$. Notice that the dimension of the group velocity is sec$^{-1}$ because we
define $V_g=\partial \omega/\partial Q$, with $Q=q/k_F$ as the reduced wave vector. One can easily notice
that the intersubband CME attains its magnetoroton shape only at $B \ge 1.0$ T for which values the $V_g$
curves cross the zero twice: the first for the maxon and the second for the roton. As such, there is a
minimum (threshold) value of $B$ (i.e., $B_{min}$) below which the MR does not exist.

\begin{figure}[htbp]
\includegraphics*[width=8cm,height=9cm]{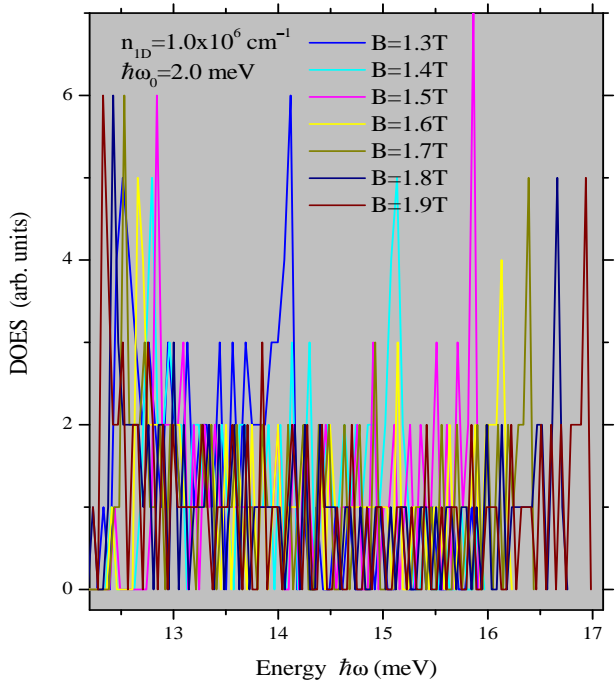}
\caption{(Color Online) The density of excitation states (DOES) of the MR excitations plotted in Fig. 1 as
a function of energy $\hbar\omega$ for several values of $B$ and for the given values of $n_{1d}$ and
$\hbar\omega_0$. (After M.S. Kushwaha, Ref. 26).}
\label{fig8}
\end{figure}

Figure 8 shows the density of excitation states (DOES) of the MR mode versus the energy for several values
of magnetic field and for the given values of $n_{1d}$ and $\hbar\omega_0$. An interesting feature
this figure dictates is that both maxon and roton are the higher density of excitation states.


\begin{figure}[htbp]
\includegraphics*[width=8cm,height=9cm]{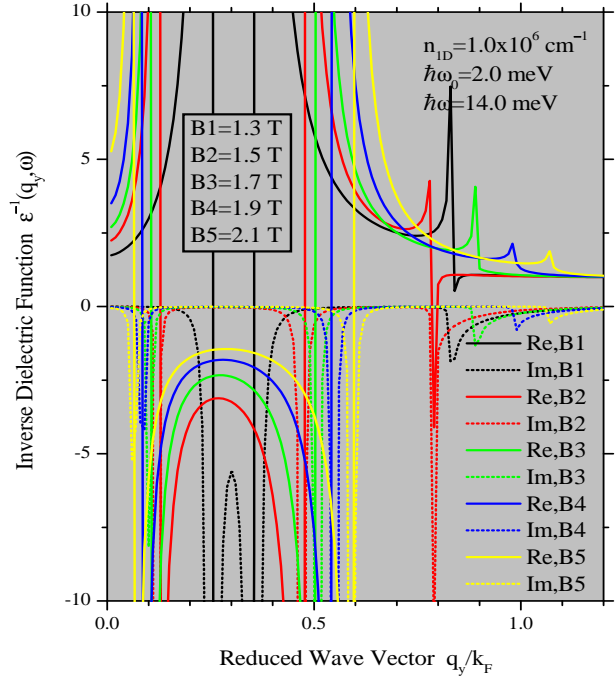}
\caption{(Color Online) Inverse dielectric function $\epsilon^{-1}(q_y, \omega)$ versus reduced propagation
vector $q_y/k_F$ for various values of the magnetic field ($B$). The other parameters are as given inside
the picture. (After M.S. Kushwaha, Ref. 27).}
\label{fig9}
\end{figure}

Figure 9 illustrates the inverse dielectric function $\epsilon^{-1}(q_y, \omega)$ as a function of the reduced
propagation vector $q_y/k_F$ for various values of the magnetic field ($B$). Notice that the excitation energy
is fixed at $\hbar\omega=14.0$ meV. It is noteworthy that the quantity that directly affects the transport
phenomenon is the spectral weight Im[$\epsilon^{-1}(q_y, \omega)$], which contains both the single-particle
contribution at large propagation vector ($q_y$) and the collective (magnetoplasmon) contribution at small
$q_y$. It is clearly observed that, for $B=0.21$ T, the sharp peaks at the reduced wave
vector $q_y/k_F\simeq 0.055$ and 0.596 stand for the two positions of the magnetoroton for the given energy $\hbar\omega=14.0$ meV: the first corresponds to the phononic part and the second to the rotonic one.
Similarly, the broader peak at $q_y/k_F\simeq 1.069$ refers to the respective intersubband single-particle
excitations (SPE). The latter actually stands for the position where the roton mode merges with the
respective intersubband SPE. The other peaks for different values of the magnetic field $B$ can be easily
identified. Clearly, these peaks are a result of the existing poles of the inverse dielectric function at the
corresponding values of energy ($\hbar\omega$) and propagation vector ($q_y$).

\begin{figure}[htbp]
\includegraphics*[width=8cm,height=9cm]{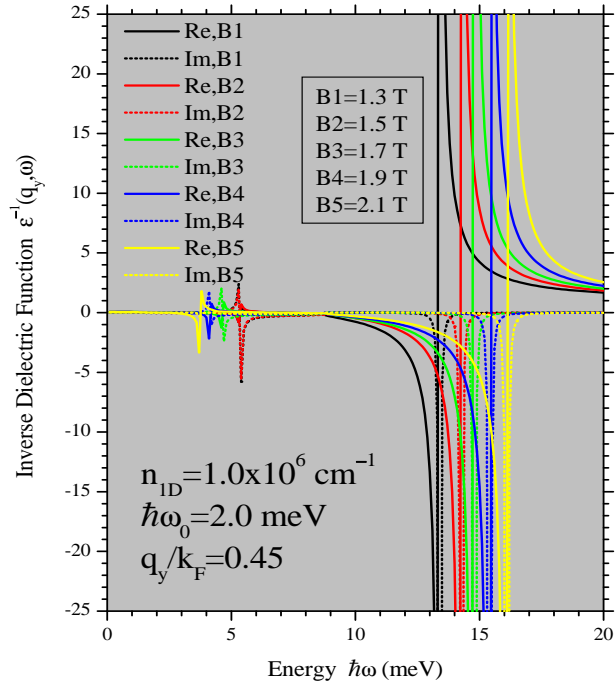}
\caption{(Color Online) Inverse dielectric function $\epsilon^{-1}(q_y, \omega)$ versus energy
$\hbar \omega$ for various values of magnetic field ($B$). Other parameters are listed inside
the picture. (After M.S. Kushwaha, Ref. 27).}
\label{fig10}
\end{figure}

Figure 10 illustrates the inverse dielectric function $\epsilon^{-1}(q_y, \omega)$ as a function of the
excitation energy $\hbar\omega$ for various values of the magnetic field ($B$). The reduced propagation
vector is fixed at $q_y/k_F=0.45$. Again, in analogy with the previous discussion, we focus on the
imaginary part of the inverse dielectric function. Since we choose a relatively small values of
the propagation vector, the peaks in this figure can only identify the collective (magnetoplasmon)
excitations. For the magnetic field $B=2.1$ T, we notice that the lower peak
at $\hbar\omega \simeq 3.737$ meV stands for the intrasubband magnetoplasmon in the quantum wire,
whereas the higher peak at $\hbar\omega \simeq 16.059$ meV corresponds to the rotonic part of the
intersubband magnetoplasmon, which is identified as the magnetoroton mode in the corresponding system.
Other peaks for different values of the magnetic field can be similarly identified. For such a smaller
value of the propagation vector, no resonance peak corresponds to any single-particle excitation
spectrum in the system.

It is interesting to notice the occurrence of the resonance peaks in both figures with the variation of
the magnetic field. Figure 9 shows that the lower (higher) peak of the magnetoroton mode occurs at lower
(higher) values of the propagation vector as the magnetic field is enhanced. Also, generally speaking,
the larger the magnetic field, the greater the propagation vector for the resonance peaks corresponding
to the single-particle excitations. Figure 10 depicts that the larger the magnetic field, the smaller
(greater) the excitation energy of the intrasubband (intersubband, or magnetoroton) mode. This leads us
to infer that the larger the magnetic field, the greater the position of the magnetoroton mode in the
($\omega, q$) space of the spectrum. A similar conclusion can be drawn by having a closer look at Fig. 4
above, where we had, in fact, searched the zeros of the dielectric function. Another interesting aspect
is that the resonance peaks associated with the magnetoroton mode are sharper and greater (in magnitude)
than those related with the intrasubband magnetoplasmons and/or the single-particle excitations. This
remark is equally valid for both Figs. 9 and 10.


\begin{figure}[htbp]
\includegraphics*[width=8cm,height=9cm]{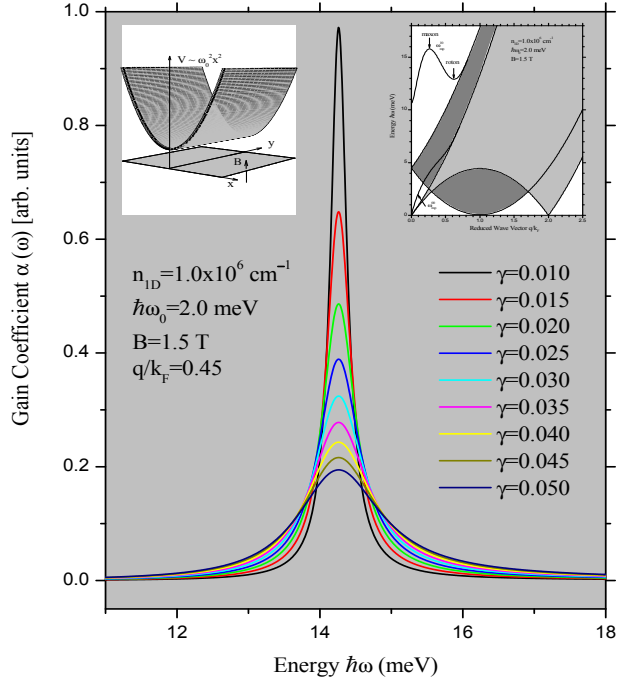}
\caption{(Color online) The gain coefficient $\alpha$($\omega$) as a function of the excitation energy
$\hbar\omega$ for several values of the damping factor and for the given values of the magnetic field $B$,
the charge density $n_{1d}$, and the confinement energy $\hbar\omega_0$. The parameters are as given inside
the picture. The inset on the top-left shows the model quantum wire investigated here. The inset on the
top-right illustrates the magnetoplasmon excitation spectrum in a two-subband model within the full RPA [8].
The crux of the matter here is the intersubband magnetoplasmon (or magnetoroton) mode which observes one
maximum [the maxon] and one minimum [the roton] (after starting at $q=0$ and $\hbar \omega =10.65$ meV)
before merging with the respective single-particle continuum. (After M.S. Kushwaha, Ref. 28).}
\label{fig11}
\end{figure}

Figure 11 illustrates the computation of the gain coefficient $\alpha$($\omega$) as a function of the excitation
energy $\hbar\omega$ for various values of the damping factor $\gamma$. The gain coefficient in the context of
the laser amplification is defined as $\alpha$($\omega$)= ($\omega/2c$) Im [$\epsilon^{-1}_{inter} (q,\omega)$];
$c$ being the speed of light in vacuum. The symbol Im refers to the imaginary part of nonlocal, dynamic, inverse
dielectric function (considered only for the relevant intersubband [or magnetoroton] excitations). The gain
coefficient that persists due to the electronic transitions shows a maximum at $\hbar\omega \simeq 14.26$ meV
for the damping factor $\gamma=0.010$. It is not just by chance that the peak position occurs at the expected
values of ($q, \omega$) in the excitation spectrum. As it is intuitively expected, the peak turns towards the
lower energy with increasing damping factor. It is reasonable that an amplifier device such as a laser gain
medium cannot maintain a fixed gain for arbitrarily high input powers, because this would require adding
arbitrary amounts of power to the amplified signal. Therefore, the gain must be reduced for high input powers;
this phenomenon is called {\em gain saturation}. In the case of a laser gain medium, it is widely known that
the gain does not instantly adjust to the level according to the optical input power, because the gain medium
stores some amount of energy, and the stored energy determines the gain.

Given the sequence of instances manifesting in the system, it makes sense to reason that the applied magnetic
field drives the system to a metastable (or non-equilibrium) state [see, e.g., Fig. 12] that gives rise to the
population inversion so that gain rather than absorption occurs at the frequencies of interest. This is
attested by the existing negative group velocity (NGV) associated with the anomalous dispersion in the range
between maxon and roton.

It is evident from Fig. 11 that the lower the damping (or ohmic or scattering loss), the higher the gain in the
medium. The bandwidth of the laser amplifier is seen to becoming narrower with increasing gain. The concept of
bandwidth in the laser amplification is different from that in the band structure. Conventionally, the
bandwidth of an amplifier is defined as the full distance between the frequency (or energy) points at which the
amplifier gain has dropped to half the peak value. Another important issue is the nature of the electronic
transitions: an amplifying (absorbing) transition implies a positive (negative) values
of Im [$\epsilon^{-1}_{inter} (q,\omega)$] and hence of $\alpha$($\omega$). Of course, one can always associate
a suitable $\pm$ sign with $\alpha$($\omega$) in order to give it the proper meaning for either amplifying or
absorbing medium.

One may argue that the sign specifying the gain or loss should result from the calculus. As a matter of fact,
this is what we should expect if we are not sure of the characteristics of the medium. But if we know (as is
the case here) that NGV implies gain rather than loss in a certain frequency range [between maxon and roton],
then we do have the freedom to choose the sign of the damping term. Similar remarks as made on the sign
convention here can be seen in the biblical textbooks on lasers.

These characteristics open the possibility of designing MR-based electron device capable of amplifying a small
optical signal of definite wavelength. Figure 11 then clearly provides a platform for realizing the potential
of a magnetized quantum wire to act as an {\em active} laser medium. The situation is analogous to the
(quasi-two dimensional) superlattices where the crystal can exhibit a negative resistance: it can refrain
from consuming energy like a resistor and instead feed energy into an oscillating circuit [6].

\begin{figure}[htbp]
\includegraphics*[width=8cm,height=9cm]{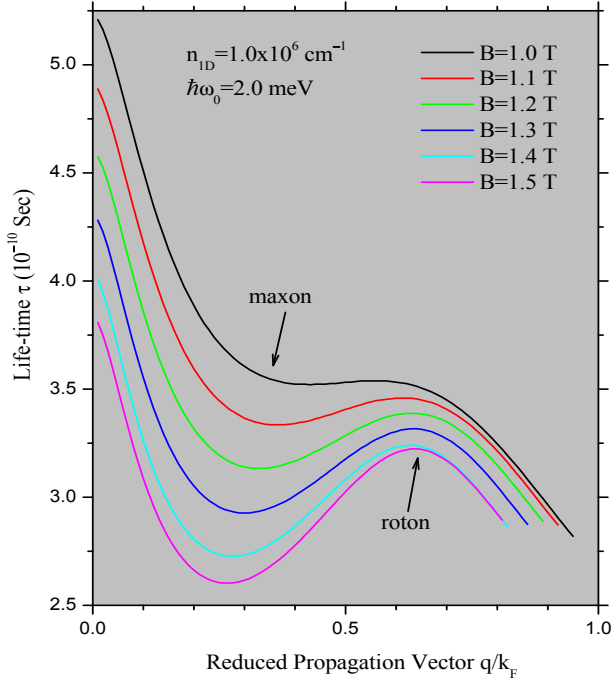}
\caption{(Color online) The life-time $\tau$ vs. the reduced propagation vector $q/k_F$ for the magnetoroton
for several values of the magnetic field $B$ [with $B \geq B_{th}$]. The other parameters are as given
inside the picture. We call attention to the trend: the stronger the magnetic field, the shorter the life-time,
and hence more susceptible the metastable state around the maxon. (After M.S. Kushwaha, Ref. 28).}
\label{fig12}
\end{figure}

Figure 12 shows the life-time $\tau$ vs. the reduced propagation vector $q/k_F$ for the magnetoroton for several
values of the magnetic field $B$. This figure brings about the fact that the applied magnetic field drives the
system to a metastable (or non-equilibrium) state. The metastable state is an important concept, from both a
fundamental and a practical point of view, in the condensed matter physics. It is, by definition, a state that
may exist even though it is much less stable than its ultimate equilibrium state. Irradiation of the system
with the light of suitable wavelength allows its electrons to jump to an excited state. When the incoming
radiation is removed, the excited electron goes back to its original (lower) state. However, when an electron
goes to a metastable state, it remains there for a relatively longer duration. This process leads to
accumulation of electrons in the metastable state, since the rate of addition of electrons to the metastable
state is higher than the rate of their de-excitation. This leads to the phenomenon called
{\em population inversion}, which forms the basis of lasing action of lasers. There are different ways to
represent the metastable state for a given system. We choose to compute the life-time of magnetoroton in the
$\omega - q$ space until it propagates as a bonafide intersubband excitation [i.e., before it merges with the
respective single-particle continuum] (see Fig. 3). Just as expected, the roton stands as an unstable
transition state between the metastable state near the maxon and the ultimate equilibrium state at lower
wavelengths. The picture speaks clearly: the stronger the magnetic field, the shorter the life-time, and hence
more vulnerable the metastable state. The damping can be interpreted as the inverse of the life-time of the
excited quantum state. This implies that the smaller magnetic fields favor the optimum case for the higher gain coefficients.

\section{Concluding Remarks}

In summary, we have gathered and reviewed recent results obatined on the fundamental aspects associated with
the intersubband collective (magnetoroton) excitations in the magnetized quantum wires within the RPA. The
only necessary and sufficient condition for the existence of magnetorotons in quantum wires
is: $B \ge B_{th}$. The crucial part of the investigation is the negative group velocity observed by the
magnetoroton between maxon maximum and roton minimum. The distinctive feature of NGV is that it leads to a
tachyon-like (superluminal) behavior without one's having to introduce negative energies. The NGV is
associated with the anomalous dispersion in a gain medium, where the sign of the phase velocity is the same
for the incident and transmitted waves and energy flows inside the gain medium in the opposite direction to
the incident energy flow in vacuum. The insight is that demonstration of NGV is possible in media with
inverted population -- an outcome of the metastable state caused by the magnetic field satisfying
$B \ge B_{th}$ -- so that gain instead of absorption occurs at the frequencies of interest. A medium with
population inversion has the remarkable ability of amplifying a small optical signal of definite wavelength;
i.e., it can serve as an {\em active} laser medium.

The MR minimum is the mode of higher density of excitation states. It is worth mentioning that roton features
are among the most significant manifestations of the many-particle interactions. They arise from the interplay
between direct and exchange terms of the electron gas and the depth of the minimum is determined by the
strength of the exchange vertex corrections. As such, incorporating the many-body effects adequately should
give a better insight into the propagation characteristics of magnetoroton.

This fundamental investigation suggests an interesting and important application: the electronic device based
on such magnetoroton modes can act as an optical amplifier. Since all the parameters [such as the charge
density, magnetic field, confinement potential, propagation vector, etc.] involved in the process that leads
us to infer this proposal are in the reach of the current technology, the concept should prompt the device
experiments. 

Ever since its discovery [32], graphene -- a one-atom thick sheet of carbon atoms arranged laterally in a
honeycomb lattice -- has drawn tremendous attention of the scientific community world wide. It is the
{\em real} thinnest material in the universe known so far, although similar synthesis of silicene [33] and
germanene [34] have recently been reported. Since it is real, it must have a {\em finite} thickness. This
allows the device physicists to electrostatically fabricate quasi-1- and quasi-0-dimensional (Dirac)
electron systems -- the equivalent of semiconducting quantum wires and quantum dots -- out of graphene.
Then it makes sense to probe graphene in the quest of magnetorotons and its role as an optical amplifier.

It is premature to predict whether the tantalizing concept of magnetized quantum wire as an optical amplifier
will emerge as an exciting idea to be engaged in by the researchers. But certainly no other system of reduced
dimensions has spoiled scientists and engineers with as many appealing features to pursue.

\vspace{1.0cm}
{\centerline {\em After I'm dead I'd rather have people ask why I have no monument than why I have one.}}
{\centerline {\hspace{13.5cm} --- Unknown}}

\newpage
\begin{acknowledgments}
During the course of this work, the author has benefited from many enlightening discussions and useful
communications with some colleagues. I would like to particularly thank Allan MacDonald, Hiroyuki
Sakaki, Loren Pfeiffer, Daniel Gammon, Aron Pinczuk, Harald Ibach, Ray Egerton, and Peter Nordlander.
I sincerely thank Professor Thomas Killian for all the support and encouragement. Finally, I would like
to express my deep appreciation to Kevin Singh for his unfailing and timely help with the software.
\end{acknowledgments}




\begin{references}
\bibitem[1]{1} J.R. Schrieffer, in {\em Semiconductor Surface Physics} edited by R.H. Kingston (University
               of Pennsylvania Press, Philadelphia, PA, 1957) p. 55.
\bibitem[2]{2} A.B. Fowler, F.F. Fang, W.E. Howard, P.J. Stiles, Phys. Rev. Lett. {\bf 16}, 901 (1966).
\bibitem[3]{3} K.v. Klitzing, G. Dorda, and M. Pepper, Phys. Rev. Lett. {\bf 45}, 494 (1980).
\bibitem[4]{4} D.C. Tsui, H.L. Stormer, and A.C. Gossard, Phys. Rev. Lett. {\bf 48}, 1559 (1982).
\bibitem[5]{5} H. Sakaki, Jpn. J. Appl. Phys. {\bf 19}, L735 (1980).
\bibitem[6]{6} For an extensive review of electronic, optical, and transport phenomena in the systems
               of reduced dimensions such as quantum wells, quantum wires, and quantum dots, see M.S.
               Kushwaha, Surf. Sci. Rep. {\bf 41}, 1 (2001).
\bibitem[7]{7} S. Tomonaga, Prog. Theoret. Phys. {\bf 5}, 544 (1950).
\bibitem[8]{8} J.M. Luttinger, J. Math. Phys. {\bf 4}, 1154 (1963).
\bibitem[9]{9} B.Y.K. Hu and S. Das Sarma,  Phys. Rev. Lett.  {\bf 68}, 1750 (1992).
\bibitem[10]{10} K. F. Berggren, T. J. Thornton, D. J. Newson, and M. Pepper, Phys. Rev. Lett.
                 {\bf 57}, 1769 (1986).
\bibitem[11]{11} G. Timp, A. M. Chang, P. Mankiewich, R. Behringer, J. E. Cunningham, T. Y. Chang, and
                 R. E. Howard, Phys. Rev. Lett. {\bf 59}, 732 (1987).
\bibitem[12]{12} M. L. Roukes, A. Scherer, S. J. Allen, Jr., H. G. Craighead, R. M. Ruthen, E. D. Beebe,
                 and J. P. Harbison, Phys. Rev. Lett. {\bf 59}, 3011 (1987).
\bibitem[13]{13} B. J. van Wees, H. van Houten, C. W. J. Beenakker, J. G. Williamson, L. P. Kouwenhoven,
                 D. van der Marel, and C. T. Foxon, Phys. Rev. Lett. {\bf 60}, 848 (1988).
\bibitem[14]{14} T. Demel, D. Heitmann, P. Grambow, and K. Ploog, Phys. Rev. Lett. {\bf 66}, 2657 (1991).
\bibitem[15]{15} A.R. Go\~ni, A. Pinczuk, J.S. Weiner, B.S. Dennis, L.N. Pfeiffer, and K.W. West,
               Phys. Rev. Lett. {\bf 70}, 1151 (1993).
\bibitem[16]{16} O. M. Auslaender, H. Steiberg, A. Yacoby, Y. Tserkovnyak, B. I. Halperin, K. W. Baldwin,
                 L. N. Pfeiffer, and K. W. West, Science {\bf 308}, 88 (2005). This is an experimental
                 proof of spin-charge separation [predicted by E. H. Leib and F. Y. Wu, Phys. Rev. Lett.
                 {\bf 20}, 1445 (1968)] observed in the ballistic, parallel quantum wires in GaAs/AlGaAs
                 heterostructure.
\bibitem[17]{17} Q. P. Li and S. Das Sarma, Phys. Rev. B {\bf 44}, 6277 (1991).
\bibitem[18]{18} S.R.E. Yang and G.C. Aers,  Phys. Rev. B {\bf 46}, 12456 (1992).
\bibitem[19]{19} L. Wendler and V. G. Grigoryan, Phys. Rev. B {\bf 49}, 13607 (1994).
\bibitem[20]{20} The Kohn theorem [W. Kohn, Phys. Rev. {\bf 123}, 1242 (1961)] was generalized for quantum
                 wells, quantum wires, and quantum dots, respectively, in: L. Brey, N. F. Johnson, and B. I.
                 Halperin, Phys. Rev. B {\bf 40}, 10647 (1989);  Q. P. Li, K. Karrai, S. K. Yip, S.
                 Das Sarma, and H. D. Drew, Phys. Rev. B {\bf 43}, 5151 (1991); and F. M. Peeters, Phys.
                 Rev. B {\bf 42}, 1486 (1990).
\bibitem[21]{21} D. Pines, {\it The Many-Body Problem} (Benjamin, New York, 1961); A. L. Fetter and J. D.
                 Walecka, {\it Quantum Theory of Many-Particle Systems} (McGraw-Hill, New York, 1971); G.
                 D. Mahan, {\it Many Particle Physics} (Plenum, New York, 1981).
\bibitem[22]{22} L.D. Landau, J. Phys. (U.S.S.R.) {\bf 5}, 71 (1941); {\bf 8}, 1 (1941).
\bibitem[23]{23} R.P. Feynman, {\it Progress in Low Temperature Physics} {\bf 1}, 17 (1955).
\bibitem[24]{24} S.M. Girvin, A.H. MacDonald, and P.M. Platzman, Phys. Rev. Lett. {\bf 54}, 581 (1985).
\bibitem[25]{25} M.S. Kushwaha, Phys. Rev. B {\bf 76}, 245315 (2007).
\bibitem[26]{26} M.S. Kushwaha, Phys. Rev. B {\bf 78}, 153306 (2008).
\bibitem[27]{27} M.S. Kushwaha, J. Appl. Phys. {\bf 106}, 066102 (2009).
\bibitem[28]{28} M.S. Kushwaha, J. Appl. Phys. {\bf 109}, 106102 (2011).
\bibitem[29]{29} S. Das Sarma and E. H. Hwang, Phys. Rev. Lett. {\bf 102}, 206412 (2009).
\bibitem[30]{30} M.S. Kushwaha and F. Garcia-Moliner, Phys. Lett. A {\bf 205}, 217 (1995).
\bibitem[31]{31} L.J. Wang, A. Kuzmich, and A. Dogariu, Nature {\bf 406}, 277 (2000).
\bibitem[32]{32} K. S. Novoselov, A. K. Geim, S. Morozov, D. Jiang, Y. Zhang, S. Dubonos, I. Grigorieva,
                 and A. A. Firsov, Science {\bf 306}, 666 (2004).
\bibitem[33]{33} P. Vogt, P. De Padova, C. Quaresima, J. Avila, E. Frantzeskakis, M.C. Asensio, A. Resta,
                 B. Ealet, and G. Le Lay, Phys. Rev. Lett. {\bf 108}, 155501 (2012).
\bibitem[34]{34} M.E. D\'{a}vila, L. Xian, S. Cahangirov, A. Rubio, and G. Le Lay, New J. Phys. {\bf 16},
                 096002 (2014).
\end{references}
\end{document}